\RequirePackage{lineno} 
\documentclass[aps,preprint]{revtex4}

\usepackage{graphicx}
\usepackage{dcolumn}
\usepackage{bm}
\usepackage{epstopdf}
\usepackage{lineno}

\begin{document}


\title{Terahertz streaking of few-femtosecond relativistic electron beams}

\author{Lingrong Zhao$^{1,2}$, Zhe Wang$^{1,2}$,  Chao Lu$^{1,2}$, Rui Wang$^{1,2}$, Cheng Hu$^{3,4}$, Peng Wang$^{5}$, Jia Qi$^{5}$, Tao Jiang$^{1,2}$, Shengguang Liu$^{1,2}$, Zhuoran Ma$^{1,2}$, Fengfeng Qi$^{1,2}$, Pengfei Zhu$^{1,2}$, Ya Cheng$^{5,6}$, Zhiwen Shi$^{3,4}$, Yanchao Shi$^{7}$, Wei Song$^{7}$, Xiaoxin Zhu$^{7}$, Jiaru Shi$^{8}$, Yingxin Wang$^{8}$, Lixin Yan$^{8}$, Liguo Zhu$^{9}$, Dao Xiang$^{1,2,10*}$ and Jie Zhang$^{1,2*}$}
\affiliation{%
$^1$ Key Laboratory for Laser Plasmas (Ministry of Education), School of Physics and Astronomy, Shanghai Jiao Tong University, Shanghai 200240, China \\
$^2$ Collaborative Innovation Center of IFSA (CICIFSA), Shanghai Jiao Tong University, Shanghai 200240, China \\
$^3$ Key Laboratory of Artificial Structures and Quantum Control (Ministry of Education), School of Physics and Astronomy, Shanghai Jiao Tong University, Shanghai 200240, China\\
$^4$ Collaborative Innovation Center of Advanced Microstructures, Nanjing 210093, China\\
$^5$ State Key Laboratory of High Field Laser Physics, Shanghai Institute of Optics and Fine Mechanics, Chinese Academy of Sciences, Shanghai 201800, China\\
$^6$ State Key Laboratory of Precision Spectroscopy, East China Normal University, Shanghai 200062, China\\
$^7$ Science and Technology on High Power Microwave Laboratory, Northwest Institute of Nuclear Technology, Xi'an, Shanxi 710024, China\\
$^8$ Department of Engineering Physics, Tsinghua University, Beijing 100084, China\\
$^9$ Institute of Fluid Physics, China Academy of Engineering Physics, Mianyang, Sichuan 621900, China\\
$^{10}$ Tsung-Dao Lee Institute, Shanghai 200240, China\\
}
\date{\today}

\begin{abstract}
Streaking of photoelectrons with optical lasers has been widely used for temporal characterization of attosecond extreme ultraviolet pulses. Recently, this technique has been adapted to characterize femtosecond x-ray pulses in free-electron lasers with the streaking imprinted by far-infrared and Terahertz (THz) pulses. Here, we report successful implementation of THz streaking for time-stamping of an ultrashort relativistic electron beam of which the energy is several orders of magnitude higher than photoelectrons. Such ability is especially important for MeV ultrafast electron diffraction (UED) applications where electron beams with a few femtosecond pulse width may be obtained with longitudinal compression while the arrival time may fluctuate at a much larger time scale. Using this laser-driven THz streaking technique, the arrival time of an ultrashort electron beam with 6 fs (rms) pulse width has been determined with 1.5 fs (rms) accuracy. Furthermore, we have proposed and demonstrated a non-invasive method for correction of the timing jitter with femtosecond accuracy through measurement of the compressed beam energy, which may allow one to advance UED towards sub-10 fs frontier far beyond the $\sim$100 fs (rms) jitter. 
\end{abstract}

\maketitle

\section{Introduction}

Ultrafast phenomena are typically studied with a pump-probe technique in which the dynamics are initiated by a pump laser and then probed by a delayed pulse \cite{FK}. Because of the \AA ngstrom-scale wavelength, both electrons and x-rays have been used as the probe pulses for watching atoms in motion during structural changes \cite{Al, YBCO, I2}. With the advent of ultrashort lasers, the temporal resolution in such experiments depends primarily on the pulse width and arrival time jitter of the probe pulse. Currently the brightest hard x-ray pulse is provided by free-electron lasers (FELs \cite{LCLS, SACLA, PAL}) and with tremendous efforts devoted it is now possible to produce sub-femtosecond x-ray pulse \cite{TS3, Ding} with its arrival time determined with femtosecond precision \cite{EOtiming, TS2, Coffee1, Coffee2}. In contrast, for electron probes as in ultrafast electron diffraction (UED \cite{UED1, UED2}), the long standing goal to deliver few femtosecond high-brightness electron beam with well-characterized arrival time still remains quite challenging.

In UED the shortest electron pulse width is mainly limited by Coulomb repulsion \cite{JAP}. In the past two decades, many methods have been developed to mitigate this effect, e.g. reducing the beam propagation length \cite{Al, Zewail}, reducing the number of electrons per pulse \cite{SEUED1, SEUED2, SEUED3}, increasing the electron beam energy \cite{UED3, UCLA, THU, OSAKA, SJTU, BNL, SLAC, DESY}, and compressing the beam with an radio-frequency (rf) buncher \cite{CPRL, CUCLA}. Among all these approaches, bunch compression provided the most significant advance that enabled a new class of experiments \cite{Gao2013, Siwick2014}. In this method, the elongated electron beam is first sent through an rf buncher cavity where the bunch head is decelerated to lower energy while the bunch tail is accelerated to higher energy; after passing through a drift, the bunch tail with higher energy will catch up with the bunch head, leading to longitudinally compressed isolated bunch. When the rf buncher cavity is replaced by a laser, the energy modulation at optical wavelength may lead to formation of attosecond electron bunch trains that can also be applied to a certain type of experiments \cite{AB1, AB2}. 

While the rf buncher technique has been widely used in UED community and very recently a relativistic electron beam as short as 7 fs (rms) has been produced \cite{CUCLA}, it is also realized that the space charge force induced pulse broadening was solved at the cost of increasing timing jitter. This is because the phase jitter in the rf cavity leads to beam energy jitter which is further converted into timing jitter at the sample \cite{SEUED2, RFC1}. Similar jitter sources exist for FELs as well \cite{EOtiming, TS2, Coffee1, Coffee2}. Such timing jitter, if not measured and corrected, will limit the temporal resolution in pump-probe applications to a similar level. Unfortunately, time-stamping techniques developed for high-charge GeV and low-energy keV electron beams can not be easily implemented for MeV UED beams. For instance, the arrival time of a GeV beam has been measured with electro-optic sampling (EOS) technique \cite{EOtiming, EOS2}, but it is difficult to reach high temporal resolution when applying this technique to a MeV UED facility where the beam charge is relatively low. Time stamping of keV electron beams with a laser triggered streak camera has been used to correct timing jitter, but the accuracy is on the order of tens of femtoseconds \cite{RFC3}. Very recently, it has been shown that a buncher cavity powered by a laser-driven THz pulse may allow one to compress a keV beam without introducing such jitter \cite{THzbuncher}. However, it requires very intense THz source in order to apply this scheme to MeV electron beam. It has been demonstrated recently that a MeV electron beam can also interact with a THz pulse effectively through the inverse FEL mechanism for beam acceleration and manipulation, but this scheme requires a dedicated undulator and careful matching of the phase and group velocity of the THz pulse with the electron beam \cite{THzUCLA}. It should also be noted that few femtosecond relativistic electron beam with intrinsically small timing jitter has also been produced in a laser wakefield accelerator where the accelerating gradient is several orders of magnitude higher than that achieved in rf guns, but the beam quality and stability still need significant improvements in order to apply the beam for UED applications \cite{LPA}.  

Here, we demonstrate a laser-driven THz streaking technique with which the arrival time of a 6 fs (rms) ultrashort electron beam was determined with 1.5 fs (rms) accuracy. With the newly developed rf deflector that provides about 2.5 fs (rms) temporal resolution, the rf buncher cavity is optimized to compress the relativistic electron beam from about 200 fs to well below 10 fs. With a narrow slit to both enhance the local THz field strength and reduce the electron beam intrinsic angular fluctuation, accurate measurement of the relativistic electron beam arrival time is achieved with a THz pulse with moderate field strength ($\sim$100 kV/cm). Because the THz pulse used for time-stamping is tightly synchronized with the laser, the measured timing information can be directly used for machine optimizations and for correcting timing jitter in laser-pump electron-probe applications. Furthermore, a non-invasive method for correcting the timing jitter of a compressed beam through measurement of the compressed beam energy has been proposed and demonstrated. This non-invasive time-stamping method is easy to implement and can be applied to both keV and MeV UED to significantly improve the temporal resolution to potentially sub-10 fs regime.

\section{THz streaking deflectogram}

The set-up for THz streaking of a few-femtosecond relativistic electron beam is shown in Fig.~1. A $\sim$50 fs (FWHM) Ti:sapphire laser at 800 nm is first split into two pulses with a 50\%-50\% beam splitter (BS1). One pulse is frequency tripled to produce electron beam in a 1.5 cell S-band (2856 MHz) photocathode rf gun. The other pulse is further split into two parts with a 10\%-90\% beam splitter (BS2) with the main pulse ($\sim$2 mJ) used to produce THz radiation through optical rectification in LiNbO$_3$ crystal \cite{TPFP} and the remaining part for \textit{in situ} characterization of the THz pulse at the interaction region through EOS technique. The $\sim3.4$ MeV electron beam is compressed by a C-band (5712 MHz) rf buncher cavity and the electron beam arrival time is measured with THz streaking in a narrow slit. In this experiment, both the THz and electron beam are running at 50 Hz. The electron beam charge is measured to be about 30 fC with a Faraday cup. 

    \begin{figure}[b]
    \includegraphics[width = 0.95\textwidth]{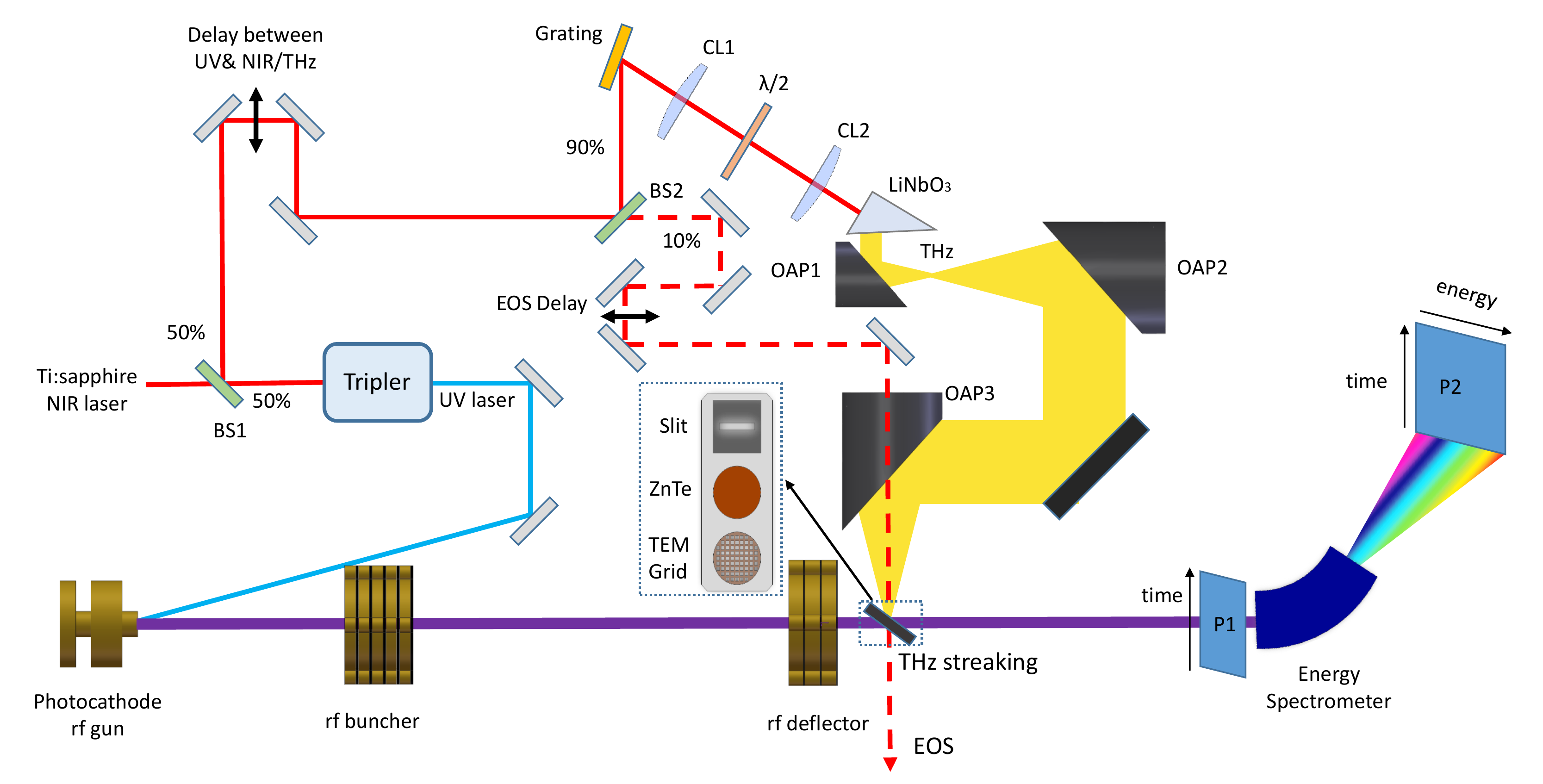}
            \caption{THz-streaking of a relativistic electron beam experiment setup. The electron beam is produced in a photocathode rf gun by illuminating the cathode with a UV laser and longitudinally compressed with an rf buncher by imprinting a negative energy chirp (i.e. with the bunch head having lower energy than the bunch tail) in the beam phase space. A set of off-axis parabolic (OAP) mirrors allows tight focus of the THz pulse onto the slit. The electron beam experiences transverse Lorenz force when passing through the slit and the THz-induced angular deflection is converted into spatial shift at the screen P1 after a drift of 1.8 m. In general, the electron beam is streaked in vertical direction with its time information mapped into spatial distribution on screen P1. Alternatively, the streaked electron beam may be sent through an energy spectrometer for measuring the longitudinal phase space at screen P2. The slit for streaking, Zinc Telluride (ZnTe) crystal for EOS, and a transmission electron microscope (TEM) grid for synchronization, are mounted on a remote-controlled manipulator.  
    \label{Fig.1}}
    \end{figure}

The laser-driven THz streaking measurement combines the standard streak camera technique with the concept of laser streaking from attosecond metrology (see, e.g. \cite{AS1, Splitring}). In this experiment, a 250 $\mu$m $\times$ 10 $\mu$m slit perforated on a 50 $\mu$m thick Al foil with laser machining is illuminated with a THz pulse with 30 degrees angle of incidence (Fig.~2). The electron beam gets a time-dependent angular kick from the THz pulse when it passes through the slit. For electrons that impinge the foil, due to multiple scattering, most of them are lost because their angles are larger than the acceptance of the vacuum pipe and a small fraction may arrive at the detector, forming a relatively uniform background. The measurement also resembles the streaking technique used to measure x-ray pulse width and timing jitter in FELs \cite{TS3, TS2, TS1}, except that here the electron beam angular distribution rather than energy distribution is changed by interaction with the THz pulse and the electron beam energy is several orders of magnitude higher than that of the photoelectrons. Very recently, such scheme has also been applied to keV energy electrons for characterizing the electron pulse width and arrival time \cite{THzbuncher}. 

    \begin{figure}[b]
    \includegraphics[width = 0.5\textwidth]{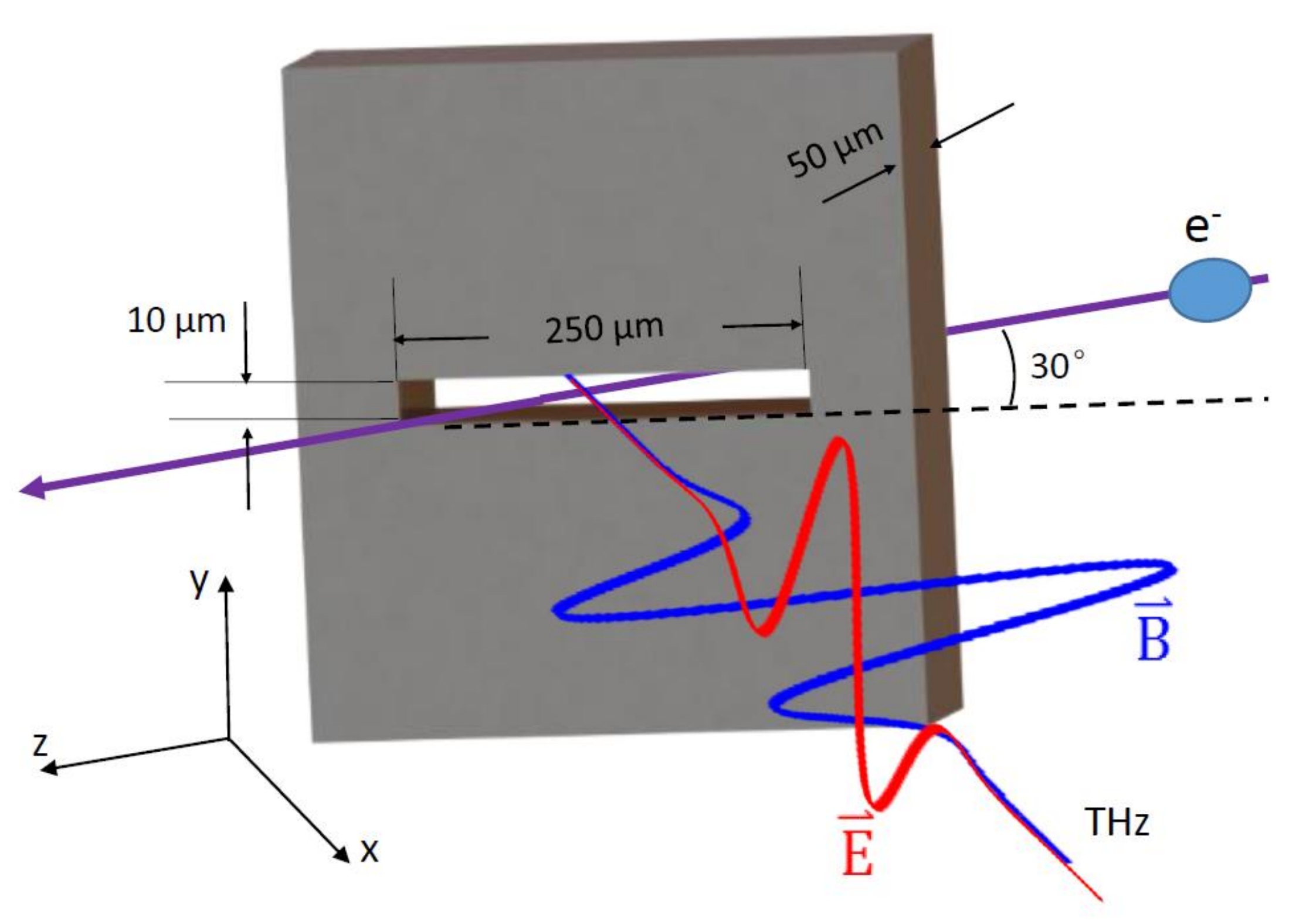}
            \caption{Schematic of the THz-electron interaction at the narrow slit.    
    \label{Fig.2}}
    \end{figure}

The THz pulse is produced with a $\sim$2 mJ 800 nm femtosecond laser through optical rectification in LiNbO$_3$ crystal with the tilted-pulse-front-pumping scheme \cite{TPFP} where the pulse front of the laser is first tilted with a diffraction grating and then imaged onto the LiNbO$_3$ with two cylindrical lenses (CL1 and CL2 in Fig.~1) for matching the laser phase velocity with the THz group velocity. The THz energy is measured to be about 0.5 $\mu$J with a calibrated Golay cell detector. With a THz camera (IRXCAM-THz-384) the THz divergence at the exit of the crystal was quantified through measurement of the THz transverse size at various positions along the propagation direction. Then a focusing system that consists of 3 off-axis parabolic mirrors are used to guide and focus the THz to the slit. The THz transverse size at the slit is measured to be about 0.5 mm (rms). The THz waveform measured by EOS \cite{EOS1} is shown in Fig.~3a with its corresponding spectrum shown in Fig.~3b. The THz field strength is calculated using the known thickness and electro-optic coefficients of ZnTe. It should be pointed out that due to crystal defects and imperfect crystal orientation, the calculated peak value of about 90 kV/cm should be considered as the lower limit of the THz field strength. Alternatively, the upper limit of the THz peak electric field is estimated to be about 170 kV/cm using the measured THz energy, transverse beam size and waveform.  

    \begin{figure}[b]
    \includegraphics[width = 0.65\textwidth]{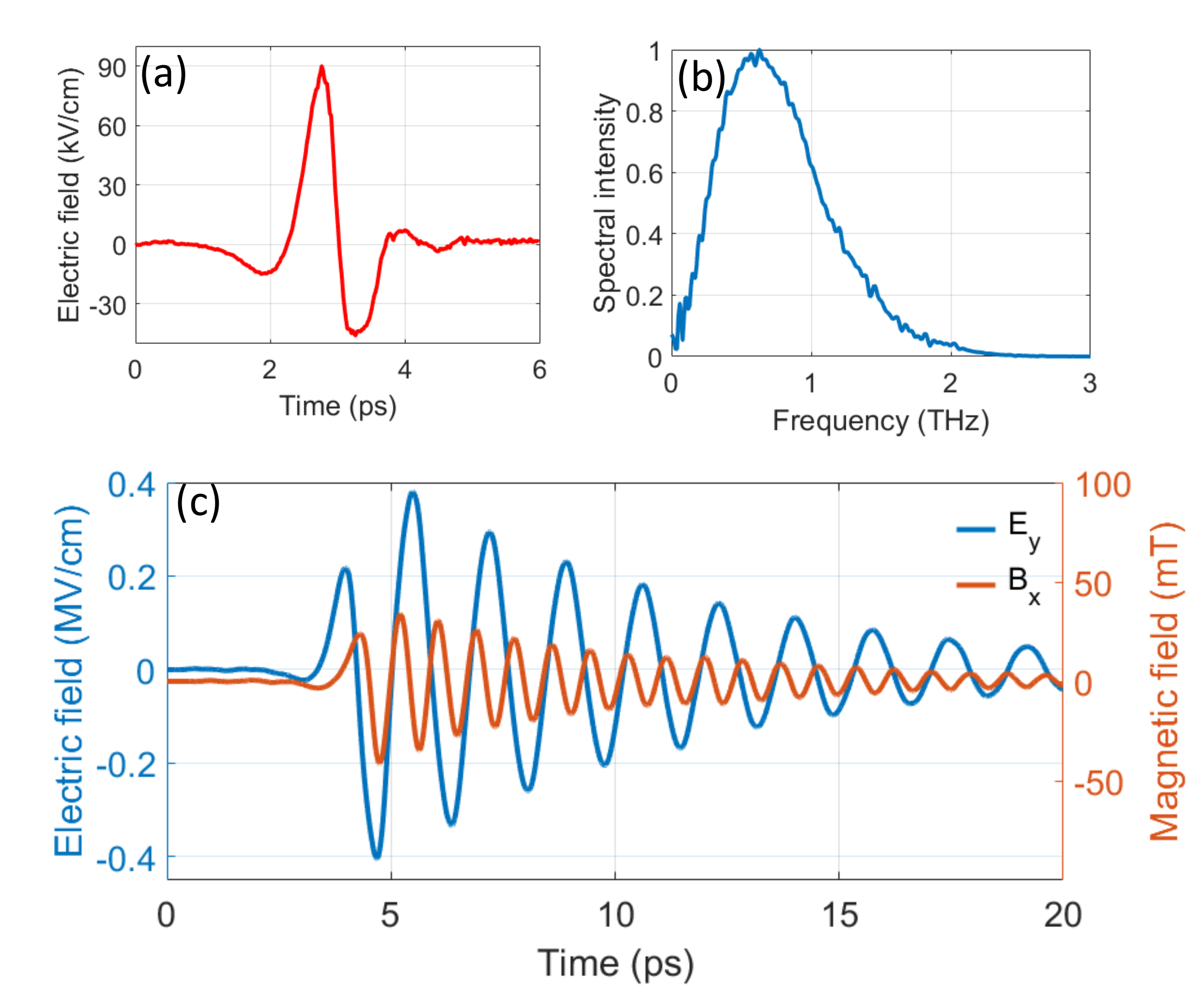}
            \caption{Measured THz waveform (a) at the streaking interaction position with EOS, the corresponding THz spectrum (b) and the simulated THz electric field and magnetic field in the center of the slit (c).     
    \label{Fig.3}}
    \end{figure}

When this single-cycle THz pulse is focused onto the narrow slit with its polarization pointing along the slit's short axis, near-field enhancement occurs in the slit that increases the streaking strength (see, e.g. \cite{TE}). The simulated field (using CST Microwave Studio) in the center of the slit is shown in Fig.~3c where one can see that due to transmission resonance, the field strength is increased by about a factor of 4 and the single-cycle THz pulse becomes multi-cycle resonating at the wavelength of approximately twice the length of the long axis of the slit, i.e. the cut off wavelength of a 250 $\mu$m $\times$ 10 $\mu$m rectangular waveguide.

This enhancement also limits the effective THz-electron interaction within the region of the slit. The simulated electron beam centroid deflection for various time delay is found with the integral of the Lorentz force along the THz-electron interaction region, as shown in Fig.~4a. Because the effective interaction region is much smaller than the wavelength of the oscillation field, the deflection (Fig.~4a) just closely follows the electric field (Fig.~3c) which dominates over the magnetic field in our interaction configuration. 
    \begin{figure}[b]
    \includegraphics[width = 0.6\textwidth]{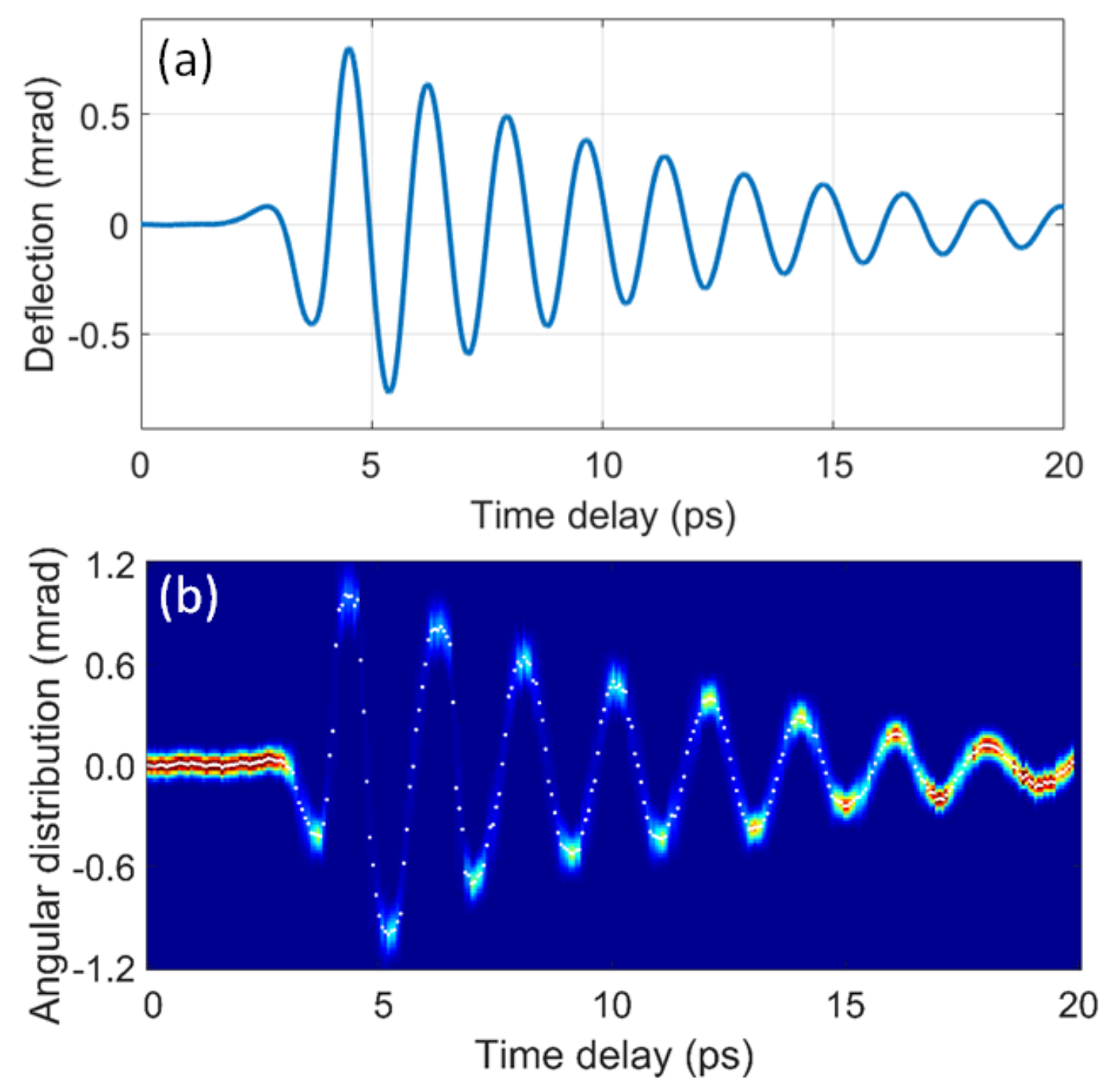}
            \caption{Simulated beam centroid deflection as a function of time delay between the electron beam and THz pulse (a) and the measured streaking deflectogram as a function of time delay (in 67 fs steps) between the electron beam and THz pulse. The measured beam centroid deflection at each time delay is shown with white points. In the measurement a collimator (3 mm upstream of the slit) is used to reduce the beam size to about 20 micron (full width) at the slit such that the electron beam feels uniform streaking force and each time slice is integrated over 50 single-shot measurements.    
    \label{Fig.4}}
    \end{figure}

Effective interaction between the THz pulse and electron beam is achieved when electron and THz beam overlap both spatially and temporally in the slit. This is done with the help of the 800 nm laser used for EOS. First, the ZnTe crystal is put in the center of the interaction chamber and the EOS signal is maximized when the 800 nm laser is well overlapped with the THz pulse both in space and time. Then the ZnTe crystal is removed from the beam path and a TEM grid is inserted. The BS2 is replaced with a mirror so that the energy of 800 nm laser is sufficient to produce transient plasmas around the interaction point on the TEM grid. The time of the laser and THz pulse is varied with a delay stage until considerable perturbation to the electron beam transverse profile from the transient electromagnetic field associated with the transient plasma is observed (see, e.g. \cite{TP1, TP2, TP3}). Note, the delay stage does not change the relative timing between the laser and THz pulse and with this technique temporal overlap between the electron beam and the THz pulse is achieved. The TEM grid is then removed from the beam path and the narrow slit is inserted. The position of the slit is varied until the 800 nm laser passes through the slit. Finally the electron beam is steered to pass through the narrow slit and both spatial and temporal overlap between the THz pulse and electron beam is then achieved. After this procedure, the delay between the electron beam and THz beam is varied and the measured streaking deflectogram is shown in Fig.~4b, which is in good agreement with the simulation result. It should be pointed out that the simulated deflection angle is slightly smaller than the experimental result, which is probably due to the fact that in the simulation the lower limit (90 kV/cm) of the peak electric field is used.

This time-dependent angular streaking allows one to map the electron beam time information into spatial distribution on a downstream screen. Similar to THz streaking in FELs, the electron beam should overlap with the THz pulse near the zero-crossing of the deflectogram (i.e. the rate of angular change is approximately linear) for measurement of beam timing jitter and the dynamic range of the measurement is limited to half of the wavelength of the streaking. The maximal streaking ramp (around t=4 ps region in Fig.~4b) is found to be 5.1 $\mu$rad/fs. The accuracy of the arrival time measurement is mainly affected by the fluctuation of the centroid divergence of the electron beam, resulting in temporal offset in the measurement. Benefiting from the narrow slit, the shot to shot fluctuation of the beam centroid divergence is found to be about 7.6 $\mu$rad, corresponding to an uncertainty of 1.5 fs in beam arrival time determination. It should be mentioned that in principle one may rotate the slit and THz polarization by 90 degrees to imprint energy modulation in the electron beam and determine the beam arrival time by monitoring the energy change of the electron beam imprinted by the THz pulse, similar to that used in \cite{THzUCLA}. However, the accuracy will be much lower because one can't benefit from the narrow slit (the slit does not reduce the beam energy fluctuation) and the buncher cavity increases the beam energy fluctuation (the buncher cavity does not increase the beam centroid divergence fluctuation). 

\section{Time-stamping of an ultrashort relativistic electron beam}

It is worth mentioning that the electron beam temporal profile can also be retrieved from the broadening of the streaked beam angular distribution. However, with the beam intrinsic divergence being approximately 50$~\mu$rad, the temporal resolution in beam temporal profile measurement is estimated to be about 10 fs (rms), limited by the strength of the streaking field. Because the available c-band rf deflector can provide a much higher resolution, in our experiment the electron beam temporal profile is measured with the rf deflector. The rf deflector is an rf structure operating in TM11 mode which gives the beam a time-dependent angular kick (i.e. $y'\propto t$) after passing through at zero-crossing phase. The beam angular distribution is converted to spatial distribution after a drift section, and the vertical axis on the phosphor screen (P1 and P2 in Fig.~1) becomes the time axis ($y \propto t$).

Fig.~5a shows measurements of 30 fC electron beam temporal profile for various voltages of the buncher cavity, corresponding to no compression ($V_b=0$), under compression ($V_b=0.84~$MV) and full compression ($V_b=1.0~$MV). The corresponding beam longitudinal phase space (Fig.~5b) was measured at screen P2 downstream of the energy spectrometer. In this measurement, the electron beam is bent in horizontal direction in the energy spectrometer such that the horizontal axis on the phosphor screen P2 becomes the energy axis ($x \propto E$). Then the beam longitudinal phase space is mapped to the transverse distribution at screen P2. The absolute time is calibrated by scanning the rf phase and recording the vertical beam centroid motion on the screens (1 degree change in rf phase corresponds to about 0.5 ps change in time). The absolute energy is calibrated from the measured magnetic field of the energy spectrometer and the dispersion at the phosphor screen. 

      \begin{figure}[b]
    \includegraphics[width = 0.9\textwidth]{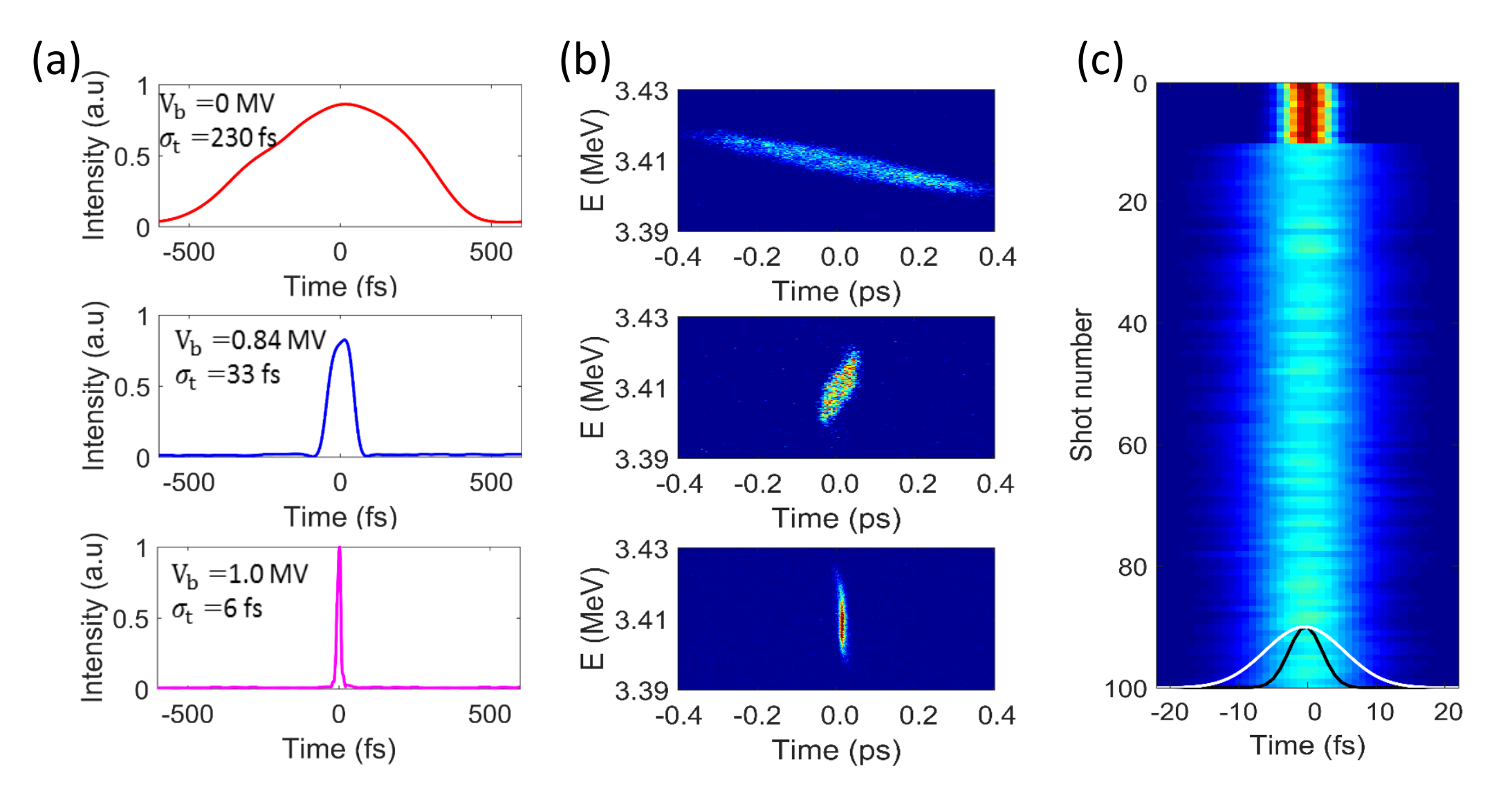}
            \caption{Longitudinal compression of a relativistic electron beam to sub-10 fs. (a) Electron beam temporal profile for various buncher voltages measured with the rf deflector. (b) Corresponding beam longitudinal phase space (bunch head to the left). (c) 100 consecutive measurements of the beam profile with rf deflector off (the first 10 shots) and on (the rest 90 shots). The average beam profiles with the deflector off (black line) and on (white line) are also shown. The number of electrons in the bunch is about $2\times10^5$. 
    \label{Fig.5}}
    \end{figure}  

As can be seen in Fig.~5b, initially the beam longitudinal phase space has positive chirp (bunch head having higher energy than bunch tail) which is caused by space charge force that accounts for bunch lengthening. As the buncher voltage is gradually increased ($V_b=0.84~$MV), the energy chirp is reversed to negative, enabling bunch compression after a drift. With the buncher voltage set to $V=1.0~$MV, the bunch tail exactly catches up with the bunch head, and the shortest bunch length was achieved. Under full compression condition, 100 consecutive measurements of the raw beam profile (with vertical axis converted into time) with rf deflector off and on are shown in Fig.~5c, where one can see that the beam has been stably compressed to about 6 fs (rms). In this measurement, the voltage of the rf deflector is about 1.8 MV and a 20 microns narrow slit is used to improve the temporal resolution of the beam temporal profile measurement to about 2.5 fs (rms), as limited by the intrinsic beam size with the rf deflector off. Because both the beam intrinsic divergence and high order effects in the rf deflector contribute to the measured vertical beam size with the deflector on \cite{CUCLA}, the estimated value of 6 fs (rms) should be considered as the upper limit of the bunch length. 

Though the bunch length is compressed to a few fs, the timing jitter is likely to be at a much larger time scale. In this particular measurement, we ignore variations in the beam temporal profile and the arrival time of the electron beam under full compression condition is determined by recording the fluctuations of the beam centroid with THz streaking. 100 consecutive measurement of beam arrival time with THz streaking is shown in Fig.~6a and the timing jitter at full compression collected over 500 shots is estimated to be about 140 fs (rms), as shown in Fig.~6b. Fortunately, such timing jitter can be corrected with femtosecond precision (as shown in Fig.~6a the jitter for most of the shots can be corrected with an accuracy better than 3 fs), which significantly improves the temporal resolution in laser-pump electron-probe applications.

      \begin{figure}[b]
    \includegraphics[width = 0.6\textwidth]{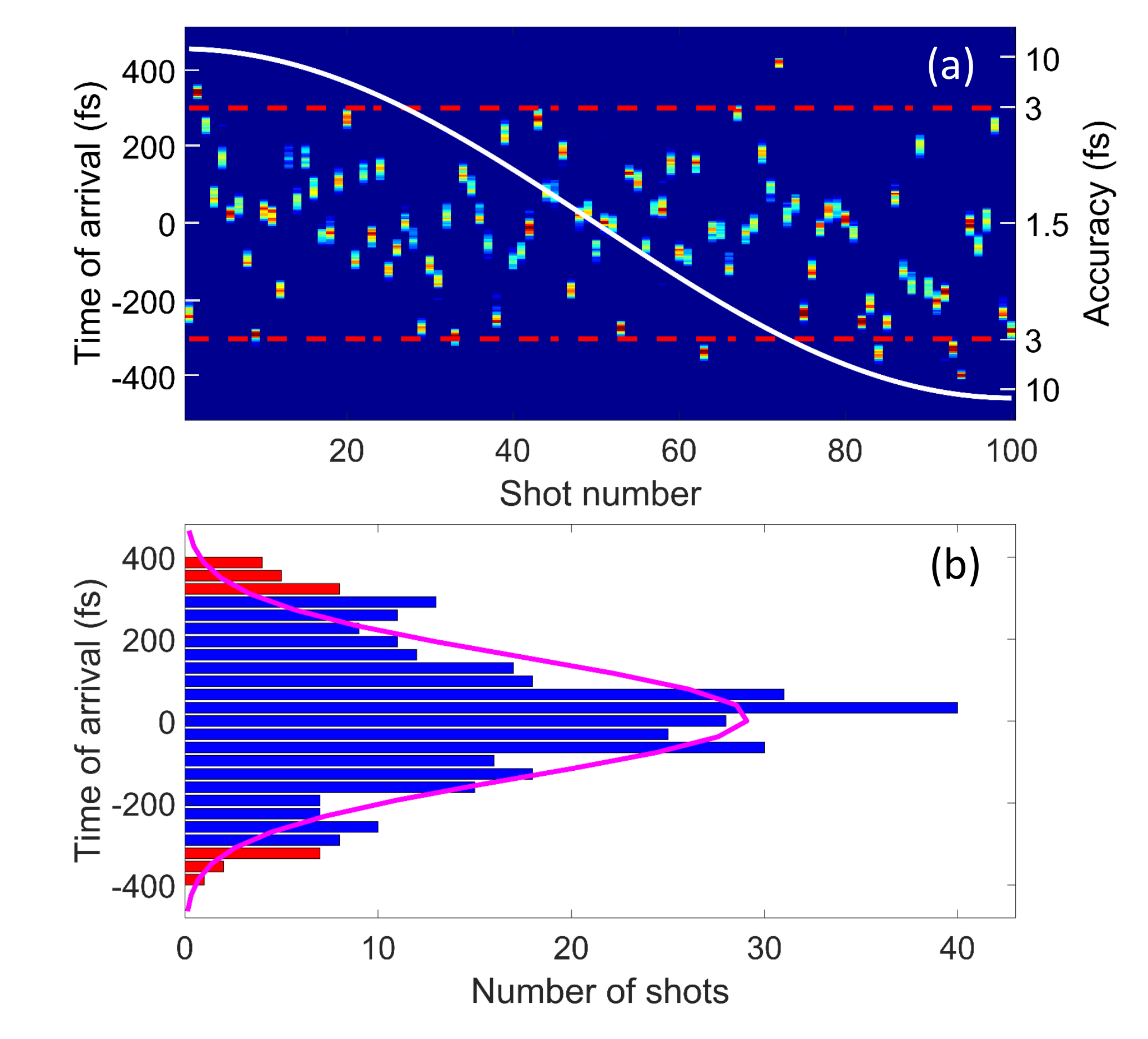}
            \caption{Time-stamping of a relativistic electron beam with THz streaking. (a) 100 consecutive measurement of beam arrival time with THz streaking using the single-valued streaking ramp (white curve). Scales on the right correspond to the accuracy in beam arrival time determination which depends on the relative time with respect to the zero-crossing of the streaking ramp. The arrival time of the shots within $\pm300~$ fs is determined with an accuracy higher than half the bunch length, i.e. 3 fs (rms). (b) Distribution of the electron beam arrival time collected over 500 shots. A Gaussian fit (magenta line) to the distribution within $\pm300~$fs yields a timing jitter of about 140 fs (rms) between the electron beam and THz pulse. The distribution with red color has uncertainty larger than 3 fs and is not used in fitting.  
    \label{Fig.6}}
    \end{figure}

\section{Real-time noninvasive time-stamping} 

The current set-up with a 10 micron slit is best suited for machine optimizations, because the narrow slit reduces the useful number of electrons in UED. While with a more intense THz source, a wider slit may be used, the dynamic range (about 600 fs in this experiment) may still hinder its applications to cases where the maximal arrival time difference is larger than half the period of the streaking field. Motivated by the fact that the timing jitter is primarily caused by energy variations after the buncher cavity, here we quantify their correlation and demonstrate that the timing jitter related to bunch compression may be corrected in a non-invasive way through measurement of the shot by shot beam energy fluctuation. In a separate experiment, the phase of the rf buncher was varied and the measured beam distribution on screen P2 is shown in Fig.~7a. In this measurement, the electron beam is streaked vertically by the THz pulse and is bent horizontally by the energy spectrometer. It should be noted that changing buncher phase is equivalent to scanning the delay time between the THz pulse and electron beam. This is the main reason that the energy-deflection map in Fig.~7a is quite similar to the time-deflection map in Fig.~4b. Combining the time information from Fig.~4b and energy information from Fig.~7a, the correlation between the arrival time and centroid energy of the beam is shown in Fig.~7b where one can see that the beam timing jitter $\Delta t$ is indeed linearly correlated with the beam energy jitter $\Delta E/E$, i.e. $\Delta t$=$R\times \Delta E/E$ with $R$ determined to be -117 ps, in good agreement with the momentum compaction of the drift \cite{RMP}, i.e. $R_{56}=-L/c\gamma^2 \approx~-120~$ps with $L\approx 1.6~$m being the distance from the buncher cavity to the THz slit and $\gamma$ being the Lorentz factor of the electron beam. 

      \begin{figure}[t]
    \includegraphics[width = 0.6\textwidth]{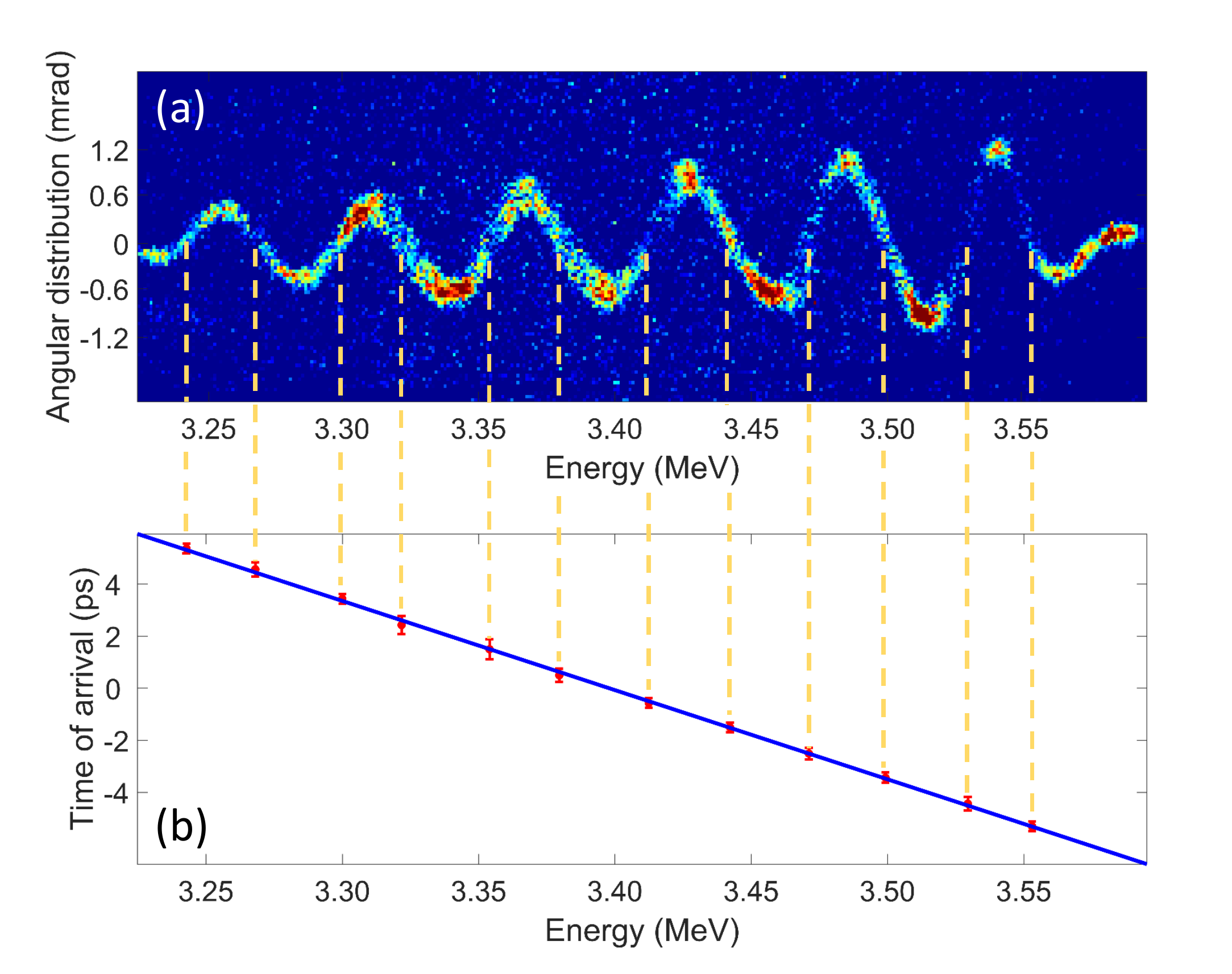}
            \caption{Energy-time correlation map. (a) Correlation between electron beam energy and beam angular distribution measured by scanning the phase of the rf buncher cavity (voltage set to $V_b=0.5~$MV for minimizing beam energy spread). (b) Beam arrival time vs beam energy after converting the angular shift into time delay. To increase the accuracy, only the  equivalent time delay for each zero-crossing points (position indicated by dashed line) of the streaking ramp is used. The data are linearly correlated and the coefficient is found to be in good agreement with the momentum compaction of the drift from the rf buncher to the THz streaking position.
    \label{Fig.7}}
    \end{figure}   

With the coefficient determined with THz streaking, the timing jitter may be corrected in a non-invasive way (e.g. using the un-diffracted beam) and the dynamical range of the jitter measurement is no longer limited by the wavelength of the THz pulse and thus even picosecond jitter may be corrected with femtosecond precision. For this method, the accuracy is limited by the uncertainty of beam energy at the entrance to the buncher cavity. In our experiment, with the beam energy stability to be about 0.02\% and the distance between the cathode to the buncher being 0.8 m (corresponding to a momentum compaction of about -60 ps), the accuracy is estimated to be about 12 fs. Note, for keV UED where the beam energy stability at the entrance to the buncher cavity is orders of magnitude higher, the accuracy of using beam energy jitter to correct timing jitter should be well below 10 fs.

\section{Conclusions and outlook}

In conclusion, we have experimentally demonstrated a novel method for time-stamping of relativistic electron beams. A non-invasive, easy-to-implement method for correcting timing jitter with high accuracy through measurement of the un-diffracted electron beam centroid energy has also been proposed and demonstrated. Together with the available few-cycle optical lasers for exciting the dynamics, the demonstrated technique should allow one to advance UED towards sub-10 fs frontier. In the future, the rf buncher voltage may be increased to produce sub-femtosecond beam. With stronger streaking field (note, LiNbO$_3$ based THz pulse with electric field exceeding 1 MV/cm has been achieved \cite{MV}) the resolution of the demonstrated method may also be extended to well beyond sub-femtosecond, making attosecond electron diffraction metrologies capable of visualizing attosecond structural dynamics within reach. 

\section{Acknowledgments}
The authors want to thank S. Li, Z. Tian and X. Su for help in THz source design. This work was supported by the Major State Basic Research Development Program of China (Grants No. 2015CB859700) and by the National Natural Science Foundation of China (Grants No. 11327902, 11504232，11655002 and 11721091). One of the authors (DX) would like to thank the support of grant from the office of Science and Technology, Shanghai Municipal Government (No. 16DZ2260200).\\
* dxiang@sjtu.edu.cn \\
* jzhang1@sjtu.edu.cn

\pagebreak


\begin{thebibliography}{30}

\bibitem{FK} F. Krausz, \textit{From femtochemistry to attophysics}, Phys. World 14, 41 (2001).

\bibitem{Al} B. Siwick, J. Dwyer, R. Jordan and R. Miller, \textit{An atomic-level view of melting using femtosecond electron diffraction}, Science 302, 1382 (2003).

\bibitem{YBCO} R. Mankowsky \textit{et al}., \textit{Nonlinear lattice dynamics as a basis for enhanced superconductivity in YBa2Cu3O6.5}, Nature 516, 71 (2014).

\bibitem{I2} J. Yang \textit{et al}., \textit{Diffractive imaging of coherent nuclear motion in isolated molecules}, Phys. Rev. Lett. 117, 153002 (2016).

\bibitem{LCLS} P. Emma \textit{et al}., \textit{First lasing and operation of an angstrom-wavelength free-electron laser}, Nat. Photonics 4, 641 (2010).

\bibitem{SACLA} T. Ishikawa \textit{et al}., \textit{A compact X-ray free-electron laser emitting in the sub-angstrom region}, Nat. Photonics 6, 540 (2012).

\bibitem{PAL} H. Kang \textit{et al}., \textit{Hard X-ray free-electron laser with femtosecond-scale timing jitter}, Nat. Photonics 11, 708 (2017).

\bibitem{TS3} W. Helml \textit{et al}., \textit{Measuring the temporal structure of few-femtosecond free-electron laser X-ray pulses directly in the time domain}, Nat. Photonics 8, 950 (2014).

\bibitem{Ding} S. Huang \textit{et al}., \textit{Generating single-spike hard X-Ray pulses with nonlinear bunch compression in free-electron lasers}, Phys. Rev. Lett. 119, 154801 (2017).

\bibitem{EOtiming} F. Tavella, N. Stojanovic, G. Geloni, and M. Gensch, \textit{Few-femtosecond timing at fourth-generation X-ray light sources}, Nat. Photonics 5, 162 (2011).

\bibitem{TS2} I. Grguras \textit{et al}., \textit{Ultrafast X-ray pulse characterization at free-electron lasers}, Nat. Photonics 6, 852 (2012).

\bibitem{Coffee1} M. Harmand \textit{et al}., \textit{Achieving few-femtosecond time-sorting at hard X-ray free-electron lasers}, Nat. Photonics 7, 215 (2013).

\bibitem{Coffee2} N. Hartmann \textit{et al}., \textit{Sub-femtosecond precision measurement of relative X-ray arrival time for free-electron lasers}, Nat. Photonics 8, 706 (2014).

\bibitem{UED1} M. Chergui and A. Zewail, \textit{Electron and x-ray methods of ultrafast structural dynamics: advances and applications}, ChemPhysChem. 10, 28 (2009).

\bibitem{UED2} G. Sciaini and R. Miller, \textit{Femtosecond electron diffraction: heralding the era of atomically resolved dynamics}, Rep. Prog. Phys. 74, 096101 (2011).

\bibitem{JAP} B. Siwick, J. Dwyer, R. Jordan, and R. Miller, \textit{Ultrafast electron optics: propagation dynamics of femtosecond electron packets}, J. Appl. Phys. 92, 1643 (2002).

\bibitem{Zewail} P. Baum, D. Yang, and A. Zewail, \textit{4D Visualization of Transitional Structures in Phase Transformations by Electron Diffraction}, Science 318, 788 (2007).

\bibitem{SEUED1} E. Fill, L. Veisz, A. Apolonski, and F. Krausz, \textit{Sub-fs electron pulses for ultrafast electron diffraction}, New J. Phys 8, 272 (2006).

\bibitem{SEUED2} L. Veisz, G. Kurkin, K. Chernov, V. Tarnetsky, A. Apolonski, F. Krausz, and E. Fill, \textit{Hybrid dc-ac electron gun for fs-electron pulse generation}, New J. Phys 9, 451 (2007).

\bibitem{SEUED3} M. Aidelsburger, F. Kirchner, F. Krausz, and P. Baum, \textit{Single-electron pulses for ultrafast diffraction}, Proc. Natl. Acad. Sci. U.S.A. 107, 19714 (2010).

\bibitem{UED3} J. Hastings, F. Rudakov, D. Dowell, J. Schmerge, J. Cardoza, J. Castro, S. Gierman, H. Loos, and P. Weber, \textit{Ultrafast time-resolved electron diffraction with megavolt electron beams}, Appl. Phys. Lett. 89, 184109 (2006).

\bibitem{UCLA} P. Musumeci, J. Moody, C. Scoby, M. Gutierrez, H. Bender, and N. Wilcox, \textit{High quality single shot diffraction patterns using ultrashort megaelectron volt electron beams from a radio frequency photoinjector}, Rev. Sci. Instrum. 81, 013306 (2010).

\bibitem{THU} R. Li \textit{et al}., \textit{Experimental demonstration of high quality MeV ultrafast electron diffraction}, Rev. Sci. Instrum. 81, 036110 (2010).

\bibitem{OSAKA} Y. Murooka, N. Naruse, S. Sakakihara, M. Ishimaru, J. Yang, and K. Tanimura,\textit{Transmission-electron diffraction by MeV electron pulses}, Appl. Phys. Lett. 98, 251903 (2011).

\bibitem{SJTU} F. Fu, S. Liu, P. Zhu, D. Xiang, J. Zhang, and J. Cao, \textit{High quality single shot ultrafast MeV electron diffraction from a photocathode radio-frequency gun}, Rev. Sci. Instrum. 85, 083701 (2014).

\bibitem{BNL} P. Zhu \textit{et al}., \textit{Femtosecond time-resolved MeV electron diffraction}, New J. Phys. 17, 063004 (2015).

\bibitem{SLAC} S. Weathersby \textit{et al}., \textit{Mega-electron-volt ultrafast electron diffraction at SLAC National Accelerator Laboratory}, Rev. Sci. Instrum. 86, 073702 (2015).

\bibitem{DESY} S. Manz \textit{et al}., \textit{Mapping atomic motions with ultrabright electrons: towards fundamental limits in space-time resolution}, Faraday Discuss. 177, 467 (2015).

\bibitem{CPRL} T. Van Oudheusden, P. Pasmans, S. Van der Geer, M. De Loos, M. Van der Wiel, and O. Luiten, \textit{Compression of subrelativistic space-charge-dominated electron bunches for single-shot femtosecond electron diffraction}, Phys. Rev. Lett. 105, 264801 (2010).

\bibitem{CUCLA} J. Maxson, D. Cesar, G. Calmasini, A. Ody, P. Musumeci, and D. Alesini, \textit{Direct measurement of sub-10 fs relativistic electron beams with ultralow emittance}, Phys. Rev. Lett. 118, 154802 (2017).

\bibitem{Gao2013} M. Gao \textit{et al}., \textit{Mapping molecular motions leading to charge delocalization with ultrabright electrons}, Nature 496, 343 (2013).

\bibitem{Siwick2014} V. Morrison, R. Chatelain, K. Tiwari, A. Hendaoui, A. Bruhacs, M. Chaker, and B. Siwick, \textit{A photoinduced metal-like phase of monoclinic VO2 revealed by ultrafast electron diffraction}, Science 346, 445 (2014).

\bibitem{AB1} Y. Morimoto and P. Baum, \textit{Diffraction and microscopy with attosecond electron pulse trains}, Nat. Phys. 14, 252 (2018).

\bibitem{AB2} M. Kozak, N. Schonenberger, and P. Hommelhoff, \textit{Ponderomotive Generation and Detection of Attosecond Free-Electron Pulse Trains}, Phys. Rev. Lett. 120, 103203 (2018).

\bibitem{RFC1} M. Gao, H. Jean-Ruel, R. Cooney, J. Stampe, M. De Jong, M. Harb, G. Sciaini, G. Moriena, and R. Miller, \textit{Full characterization of RF compressed femtosecond electron pulses using ponderomotive scattering}, \textit{Opt. Express} 20, 12048 (2012).

\bibitem{EOS2} A. Cavalieri \textit{et al}., \textit{Clocking Femtosecond X Rays}, Phys. Rev. Lett. 94, 114801 (2005).

\bibitem{RFC3} M. Gao, Y. Jiang, G. Kassier, and R. Miller, \textit{Single shot time stamping of ultrabright radio frequency compressed electron pulses}, Appl. Phys. Lett. 103, 033503 (2013).

\bibitem{THzbuncher} C. Kealhofer, W. Schneider, D. Ehberger, A. Ryabov, F. Krausz, and P. Baum, \textit{All-optical control and metrology of electron pulses}, Science 352, 429 (2016).

\bibitem{THzUCLA} E. Curry, S. Fabbri, J. Maxson, P. Musumeci, and A. Gover, \textit{Meter-scale terahertz-driven acceleration of a relativistic beam}, Phys. Rev. Lett. 120, 094801 (2018).

\bibitem{LPA} O. Lundh \textit{et al}., \textit{Few femtosecond, few kiloampere electron bunch produced by a laser–plasma accelerator}, Nat. Phys. 7, 219 (2011).

\bibitem{TPFP} J. Hebling, G. Almasi, I. Kozma, and J. Kuhl, \textit{Velocity matching by pulse front tilting for large area THz-pulse generation}, Opt. Express 10, 1161 (2002).

\bibitem{AS1} F. Krausz and M. Ivanov, \textit{Attosecond physics}, Rev. Mod. Phys. 81, 163 (2009). 

\bibitem{Splitring} J. Fabianska, G. Kassier, and T. Feurer, \textit{Split ring resonator based THz-driven electron streak camera featuring femtosecond resolution}, Sci. Rep. 4, 5645 (2014).

\bibitem{TS1} U. Fruhling \textit{et al}., \textit{Single-shot terahertz-field-driven X-ray streak camera}, Nat. Photonics 3, 523 (2009).

\bibitem{EOS1} Q. Wu and X.-C. Zhang, \textit{Free-space electro-optic sampling of terahertz beams}, Appl. Phys. Lett. 67, 3523 (1995).

\bibitem{TE} F. Garcia-Vidal, L. Martin-Moreno, T. Ebbesen, and L. Kuipers, \textit{Light passing through subwavelength apertures}. Rev. Mod. Phys. 82, 729 (2010).

\bibitem{TP1} M. Centurion, P. Reckenthaeler, S. Trushin, F. Krausz, and E. Fill, \textit{Picosecond electron deflectometry of optical-field ionized plasmas}, Nat. Photonics 2, 315 (2008).

\bibitem{TP2} P. Zhu \textit{et al}., \textit{Four-dimensional imaging of the initial stage of fast evolving plasmas}, Appl. Phys. Lett. 97, 211501 (2010).

\bibitem{TP3} C. Scoby, R. Li, E. Threlkeld, H. To, and P. Musumeci, \textit{Single-shot 35 fs temporal resolution electron shadowgraphy}, Appl. Phys. Lett. 102, 023506 (2013).

\bibitem{RMP} E. Hemsing, G. Stupakov, D. Xiang, and A. Zholents, \textit{Beam by design: Laser manipulation of electrons in modern accelerators}, Rev. Mod. Phys. 86, 897 (2014).

\bibitem{MV} H. Hirori, A. Doi, F. Blanchard, and K. Tanaka, \textit{Single-cycle terahertz pulses with amplitudes exceeding 1 MV/cm generated optical rectification in LiNbO(3)}, Appl. Phys. Lett. 99, 091106 (2011).

\end{thebibliography}
\end{document}